\documentclass[prl,twocolumn,showpacs,amsmath,amssymb]{revtex4}
\usepackage{graphicx}

\begin{document}

\title{\boldmath
Precision measurements of $B(D^+ \rightarrow \mu^+ \nu_{\mu})$, the pseudoscalar decay constant $f_{D^+}$,
and the quark mixing matrix element $|V_{\rm cd}|$\\

\vspace{0.2cm}
}
\vspace{-0.2cm}
\author{
\small
M.~Ablikim$^{1}$, M.~N.~Achasov$^{8,a}$, X.~C.~Ai$^{1}$, O.~Albayrak$^{4}$, D.~J.~Ambrose$^{41}$, F.~F.~An$^{1}$,
Q.~An$^{42}$, J.~Z.~Bai$^{1}$, R.~Baldini Ferroli$^{19A}$, Y.~Ban$^{28}$
% ...
\\
\vspace{0.2cm}
(BESIII Collaboration)\\
\vspace{0.2cm} {\it
$^{1}$ Institute of High Energy Physics, Beijing 100049, People's Republic of China\\
% ...
$^{4}$ Carnegie Mellon University, Pittsburgh, Pennsylvania 15213, USA\\
% ...
$^{8}$ G.I. Budker Institute of Nuclear Physics SB RAS (BINP), Novosibirsk 630090, Russia\\
% ...
$^{19}$ (A)INFN Laboratori Nazionali di Frascati, I-00044, Frascati, Italy; (B)INFN and University of Perugia, I-06100, Perugia, Italy\\
% ...
$^{28}$ Peking University, Beijing 100871, People's Republic of China\\
% ...
$^{41}$ University of Rochester, Rochester, New York 14627, USA\\
$^{42}$ University of Science and Technology of China, Hefei 230026, People's Republic of China\\
% ...
$^{a}$ Also at Novosibirsk State University, Novosibirsk, 630090, Russia\\
% ...
}
\vspace{0.4cm}
}
\noaffiliation

\begin{abstract}
We report a measurement of the branching fraction $B(D^+ \rightarrow
\mu^+ \nu_{\mu}) = [3.71 \pm 0.19 (\rm stat) \pm 0.06 (\rm sys)]\times
10^{-4}$ based on 2.92 ${\rm fb^{-1}}$ of data accumulated at
$\sqrt{s}=3.773$ GeV with the BESIII detector at the BEPCII collider.
This measurement, in conjunction with the Cabibbo-Kobayashi-maskawa matrix element $|V_{\rm
cd}|$ determined from a global Standard Model fit, implies a
value for the weak decay constant $f_{D^+}=(203.2 \pm 5.3 \pm 1.8)$~MeV.
Additionally, using this branching fraction measurement
together with a lattice QCD prediction for $f_{D^+}$,
we find $|V_{\rm cd}|=0.2210\pm 0.0058 \pm 0.0047$.
In either case, these are the most precise results for these quantities to date.
\end{abstract}

% PACS

\maketitle

In the Standard Model (SM) of particle physics, the $D^+$
meson can decay into $\ell^+\nu_\ell$ (where $\ell = e$, $\mu$, or $\tau$)
via annihilation mediated by a virtual $W^+$ boson.
(Throughout this paper, the inclusion of charge conjugate channels is implied.)
The decay rate depends upon the wave function overlap of
the two quarks at the origin, which is parametrized by the $D^+$ decay
constant, $f_{D^+}$.   All of the strong interaction effects between the two initial-state quarks
are absorbed into $f_{D^+}$.
In the SM, the decay width is given by~\cite{Rosner}
{\small
\begin{equation}
\Gamma(D^+ \rightarrow \ell^+\nu_{\ell})=
     \frac{G^2_F f^2_{D^+}} {8\pi}
     \mid V_{\rm cd} \mid^2
      m^2_{\ell} m_{D^+}
    \left (1- \frac{m^2_{\ell} } {m^2_{D^+}}\right )^2,
\label{eq01}
\end{equation}
}
\noindent
\hspace{-0.11cm}where $G_F$ is the Fermi coupling constant,
$V_{\rm cd}$ is the $c\to d$ Cabibbo-Kobayashi-Maskawa (CKM)
matrix element~\cite{pdg2010}, $m_{\ell}$ is the lepton mass, and
$m_{D^+}$ is the $D^+$-meson mass.

The decay constants $f_{D^+}$ and its $B^+$-meson counterpart
$f_{B^+}$ are critical parameters of heavy-flavor physics.
In $B$-meson physics, the $B^0\bar B^0$ mixing parameter
$x_B=\Delta M_{B}/\Gamma_B$ can be well measured, where $\Delta M_B$ and
$\Gamma_B$ are the mass difference between the two neutral $B$-meson
eigenstates and the mean neutral $B$-meson total width, respectively.
In the SM, assuming the CKM matrix element $|V_{\rm tb}|=1$ the $x_B$ is given by
\begin{equation}
x_B=\tau_B \frac{G^2_F M^2_{W}} {6\pi}
     \eta_B S(x_t) M_{B} f_B \sqrt{B_B} \mid V_{\rm td} \mid^2,
\label{eq_BBbmixing}
\end{equation}
\noindent
where $B_B$ is corresponding ``bag parameter" and $\eta_B S(x_t)$ is perturbatively
known~\cite{bernard}. Since $x_B$ is the theoretically and experimentally most
accessible quantity, a reliable and precise determination of $f_{B^+}$ is important for extracting
$|V_{\rm td}|$.  However, it is currently not possible to
measure $f_{B^+}$ directly from $B^+$ leptonic decays with the
required precision~\cite{b_2_tau-nu}, so, theoretical calculations of
$f_{B^+}$ have to be used in the determination of $|V_{\rm td}|$.  In
current lattice QCD (LQCD) calculations, the ratio $f_{D^+}/f_{B^+}$
is determined with a significantly better precision than the
individual quantities themselves. Thus, a precise measurement of
$f_{D^+}$ can be used to validate the LQCD calculation and
subsequently be used in conjunction with the LQCD value for
$f_{D^+}/f_{B^+}$ to make a precise estimate of $f_{B^+}$.  In turn,
the resulting $f_{B^+}$ value can be used to improve the precision of
$|V_{\rm td}|$ determined from the measured $B^0\bar B^0$ mixing
strength.

Measurements of $|V_{\rm cd}|$ have historically been based on measured branching fractions
for  semileptonic $D\to \pi\ell^+\nu_{\ell}$ decays and on measurements of charm production
cross sections in neutrino and antineutrino interactions.
However, extracting $|V_{\rm cd}|$ from exclusive semileptonic decay rates requires
a knowledge of the relevant hadronic form factor, which can have
theoretical uncertainties that are about $11\%$;
the uncertainty of $|V_{\rm cd}|$ determined from neutrino and antineutrino cross sections
is about $4.8\%$~\cite{pdg2010}.
A recent unquenched LQCD calculation of
$f_{D^+}$ claims a precision of about $2\%$~\cite{lqcd_HPQCD_UKQCD} and
provides an opportunity to improve the measured value of $|V_{\rm cd}|$
using an improved $D^+ \rightarrow \mu^+ \nu_{\mu}$ branching fraction determination.

In this paper we report measurements of
the branching fraction for $D^+\rightarrow \mu^+\nu_{\mu}$ decay
and the product of $f_{D^+}$ and $|V_{\rm cd}|$
based on analysis of 2.92 fb$^{-1}$ of data~\cite{bes3_lum} taken at
$\sqrt{s}=3.773$~GeV with the BESIII detector.
Using this measured $f_{D^+}|V_{\rm cd}|$ together with the CKM matrix element $|V_{\rm cd}|$,
we determine the pseudoscalar decay constant $f_{D^+}$. Alternatively,
using the measured $f_{D^+}|V_{\rm cd}|$ in conjunction with a lattice QCD prediction for $f_{D^+}$,
we determine the CKM matrix element $|V_{\rm cd}|$.
This more accurate determination of $|V_{\rm cd}|$ and improved
determination of $|V_{\rm td}|$ would improve the stringency of unitarity constraints
on the CKM matrix and provide an improved test of the SM.

The BESIII~\cite{bes3} detector is a cylindrical detector with a
solid-angle coverage of $93\%$ of $4\pi$ that operates at the BEPCII
$e^+e^-$ collider~\cite{bes3}.  It consists of several main
components.  A 43-layer main drift chamber (MDC) which surrounds the beam
pipe performs precise determinations of charged-particle
trajectories and provides ionization energy loss ($dE/dx$)
measurements that are used for charged-particle identification.  An
array of time-of-flight counters (TOF) is located radially outside of
the MDC and provides additional charged-particle identification
information.  The time resolution of the TOF system is 80~ps (110 ps) in the
barrel (end-cap) regions,
corresponding better than 2$\sigma$  $K/\pi$ separation for momentum below about 1 GeV/c.
The solid-angle coverage of the barrel TOF is $|\cos \theta|<0.83$,
while that of the end cap is $0.85<|\cos \theta|<0.95$,
where $\theta$ is the polar angle of the coverage.
A CsI(Tl) electromagnetic calorimeter (EMC) surrounds the
TOF and is used to measure the energies of photons and electrons.
The angular coverage of the barrel EMC is
$|\cos \theta| <0.82$. The two end caps cover
$0.83<|\cos \theta|<0.93$.
A solenoidal superconducting magnet located outside the EMC provides a 1
T magnetic field in the central tracking region of the detector.  The
iron flux return of the magnet is instrumented with 1600 m$^2$ of
resistive plate muon counters (MUC) arranged in nine layers in the barrel
and eight layers in the end caps that are used to identify muons with
momentum greater than 500~MeV/$c$.

The center-of-mass energy of $3.773$~GeV corresponds to the peak of
the $\psi(3770)$ resonance, which decays predominantly into $D\bar D$
meson pairs~\cite{pdg2010}.  In events where a $\bar D$ meson is fully
reconstructed, the remaining particles must all be decay products of
the accompanying $D$ meson.  In the following, the reconstructed meson
is called the tagged $\bar D$.  In a tagged $D^-$ data sample, events
where the recoiling $D^+$ decays to $\mu^+\nu_{\mu}$ can be cleanly
isolated and used to provide a measurement of the absolute branching
fraction $B(D^+ \to \mu^+\nu_{\mu})$.

Tagged $D^-$ mesons are reconstructed in nine decay modes:
$K^+\pi^-\pi^-$, $K^0_S\pi^-$, $K^0_S K^-$, $K^+K^-\pi^-$,
$K^+\pi^-\pi^-\pi^0$, $\pi^+\pi^-\pi^-$, $K^0_S\pi^-\pi^0$,
$K^+\pi^-\pi^-\pi^-\pi^+$, and $K^0_S\pi^-\pi^-\pi^+$.  Events that
contain at least three reconstructed charged tracks with good helix
fits and $|{\rm cos\theta} |<0.93$ are selected, where $\theta$ is the
polar angle of the charged tracks with respect to the beam direction.
All charged tracks other than those from $K^0_S$ decays are required
to have a distance of closest approach to the average $e^+e^-$
interaction point that is less than 1.0~cm in the plane perpendicular
to the beam and less than 15.0~cm along the beam direction.  These
charged tracks are then constrained to have a common vertex.  The TOF
and $dE/dx$ measurements are combined to form confidence levels for
pion ($CL_{\pi}$) and kaon ($CL_{K}$) particle identification
hypotheses. In this analysis pion (kaon) identification requires
$CL_{\pi}>CL_{K}$ ($CL_{K}>CL_{\pi}$) for tracks with momentum $p <
0.75$~GeV/$c$, and $CL_{\pi}>0.1\%$ ($CL_{K}>0.1\%$) for
$p>0.75$~GeV/$c$.

For the selection of photons from $\pi^0\to\gamma\gamma$ decays, the
deposited energy of a neutral cluster in the EMC is required to be
greater than $25$~($50$)~MeV if the crystal with the maximum deposited
energy in that cluster is in the barrel (end-cap) region~\cite{bes3}.
In addition, information about the EMC cluster hit time is used to
suppress electronic noise and energy deposits unrelated to the event.
In order to reduce backgrounds, the angle between the photon candidate
and the nearest charged track is required to be greater than
$10^{\circ}$.  A one-constraint (1C) kinematic fit is used to
constrain the invariant mass of $\gamma\gamma$ pairs to the mass of
the $\pi^0$ meson in order to reduce combinatorial backgrounds.  If the
1C kinematic fit converges with $\chi^2<100$, the pair is considered
as a candidate $\pi^0 \rightarrow \gamma\gamma$ decay.

We detect $K^0_S$ mesons that decay to a $\pi^+\pi^-$ pair.  A vertex
fit is performed on two oppositely charged tracks that are assumed to
be pions.  If the vertex fit is successful and the invariant mass of
the $\pi^+\pi^-$ is in the range between 0.485 and 0.515 GeV/$c^2$,
the $\pi^+\pi^-$ pair is taken as a candidate $K^0_S$ meson.

Tagged $D^-$ mesons are identified by their beam-energy-constrained mass $M_{\rm BC}$:
\begin{equation}
 M_{\rm BC} = \sqrt {E_{\rm beam}^2-|\vec {p}_{mKn\pi}|^2},
\end{equation}
where $m$ and $n$
($m$=0, 1, 2; $n$= 0, 1, 2, 3, or 4) denotes the numbers of kaons and
pions in the tagged $D^-$ decay mode being considered, $E_{\rm beam}$ is the beam energy,
and $|\vec p_{mKn\pi}|$ is the magnitude of the three-momentum of the $mKn\pi$ system.
In addition, the absolute value of the difference between the beam energy and the
sum of the measured energies of the $mKn\pi$ combination is required to be within
approximately $2.5\sigma_{E_{mKn\pi}}$ of zero, where $\sigma_{E_{mKn\pi}}$ is the decay-mode-dependent
standard deviation of the energy of the $mKn\pi$ system.

\begin{figure}[htbp]
\includegraphics[width=8.2cm]
{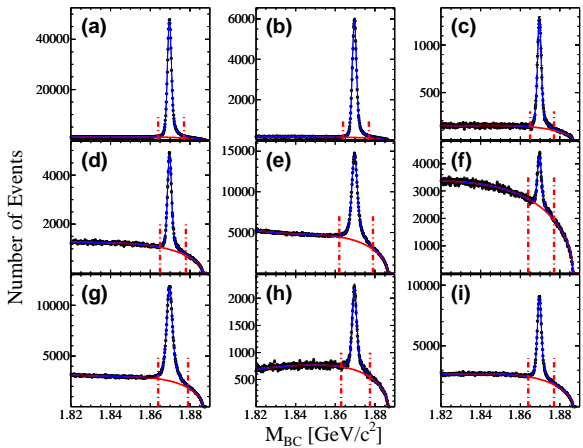}
\caption{
The beam-energy-constrained mass distributions for
the different $mKn\pi$ tagged mode combinations,
where
 (a) $K^+\pi^-\pi^-$, (b) $K^0_S\pi^-$, (c) $K^0_S K^-$, (d) $K^+K^-\pi^-$,
 (e) $K^+\pi^-\pi^-\pi^0$, (f) $\pi^+\pi^-\pi^-$, (g) $K^0_S\pi^-\pi^0$, (h) $K^+\pi^-\pi^-\pi^-\pi^+$
 and (i) $K^0_S\pi^-\pi^-\pi^+$;
the two vertical dashed red lines show the tagged $D^-$ mass region.
}
\label{fig1}
\end{figure}

The $M_{\rm BC}$ distributions for the nine $D^-$ tag modes are shown in Fig.~\ref{fig1}.
A maximum likelihood fit is used to obtain the number of tagged $D^-$ events
for each of the nine modes.
We use the Monte Carlo simulated signal shape
convolved with a double-Gaussian resolution function to represent
the beam-energy-constrained mass signal for the $D^-$ daughter particles, and
an ARGUS function \cite{Albrecht-1990am} multiplied by a third-order polynomial~\cite{bes2_D_physics_papers}
to describe the background shape to fit the $M_{\rm BC}$ distributions.
In the fits all parameters of the double-Gaussian function,
the ARGUS function and the polynomial function are left free.
We identify tagged $D^-$ candidates as combinations with $M_{\rm BC}$ within
the range given by two red dashed lines in each figure.
This requirement reduces the number of signal events by about $2\%$
and keeps a total of $1703054 \pm 3405$ tagged $D^-$~mesons ($N_{D^-_{\rm tag}}$).

Candidate $D^+ \rightarrow \mu^+\nu_{\mu}$ events are selected from
the remaining charged tracks in the system recoiling against the
tagged $D^-$-meson candidates by requiring that there be only one good
positively charged track that is identified as a $\mu^+$.  In BESIII,
a $\mu^+$ can be identified by its transit distance in the MUC,
since charged hadrons (pions or kaons) undergo strong interactions
with the absorber material and stop before penetrating very far into
the MUC.  In addition, in candidate $D^+\to\mu^+\nu_{\mu}$ events the
maximum energy $E_{\gamma\rm max}$ of any extra good photon in the EMC
is required to be less than 300 MeV.

\begin{figure}[htp]
\includegraphics[width=8.2cm]
{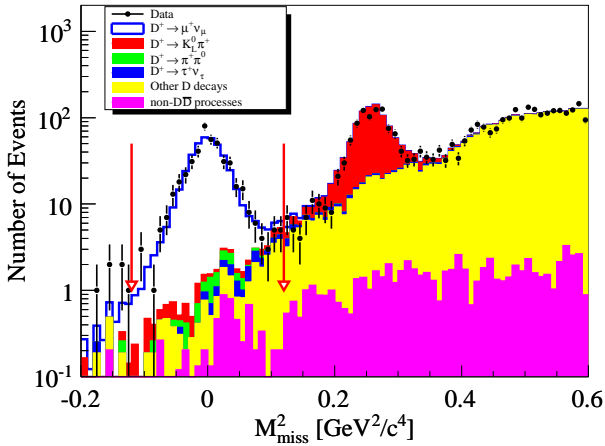}
\caption{
The $M^2_{\rm miss}$ distribution for selected single $\mu^+$
candidates,
where dots with error bars indicate the data,
the opened histogram is for Monte Carlo simulated signal events
of $D^+ \rightarrow \mu^+\nu_\mu$ decays,
and the hatched histograms are for the simulated backgrounds from
$D^+ \rightarrow K^0_L \pi^+$ (red),
$D^+ \rightarrow \pi^0\pi^+$ (green),
$D^+ \rightarrow \tau^+ \nu_{\tau}$ (blue),
all other $D$-meson decays (yellow), and
non-$D\bar D$ processes (pink).
}
\label{fig2}
\end{figure}

Since there is only a single missing neutrino in $D^+ \rightarrow
\mu^+\nu_{\mu}$ events, we require that the missing energy $E_{\rm
miss}$ and momentum $\vec p_{\rm miss}$ are such that the value of the
missing mass squared $M^2_{\rm miss}$ is consistent with zero, where
$M^2_{\rm miss}$ is defined as
\begin{equation}
M^2_{\rm miss} = (E_{\rm beam}-E_{\mu^+})^2 - (- \vec p_{D^-_{\rm tag}}- \vec p_{\mu^+} )^2.
\label{eq_miss2}
\end{equation}
Here $E_{\mu^+}$ and $\vec p_{\mu^+}$ are the energy
and three-momentum of the $\mu^+$, respectively,
and $\vec p_{D^-_{\rm tag}}$ is the three-momentum of the tagged $D^-$ candidate.
Figure~\ref{fig2} shows the $M^2_{\rm miss}$ distribution for selected
single $\mu^+$ candidates.
There are $451$ candidate $D^+ \rightarrow \mu^+\nu_{\mu}$ events
in the $|M^2_{\rm miss}|<0.12$ GeV$^2$/$c^4$ signal region as shown with two
red arrows.
The events that peak near $M^2_{\rm miss}\simeq 0.25$~GeV$^2$/$c^4$
are primarily from $D^+ \rightarrow K_L^0\pi^+$ decays, where the
$K_L^0$ is undetected.

To check the Monte Carlo simulation, we compare the $M^2_{\rm miss}$
distribution for $D^+ \to K_S^0\pi^+$ from the data with that from
Monte Carlo simulated events, where the $K_S^0$ is missing in the
calculation of $M^2_{\rm miss}$.  We select $D^+ \to K_S^0\pi^+$
events with the same requirements as these used in selection of $D^+
\rightarrow \mu^+\nu_{\mu}$, but require an additional $K^0_{S}$.  We
find that the $M^2_{\rm miss}$ resolution
for the data to be 1.194 times wider than that for the
simulated events.  To account for this difference, we scale the
$M^2_{\rm miss}$ resolution of simulated events by a factor of
$1.194$ when looking for $D^+ \to \mu^+\nu_{\mu}$ signal and
estimating numbers of peaking background events, such as $D^+ \to
K_L^0\pi^+$ and $D^+ \to \pi^+\pi^0$ decays (see below and see Fig.~\ref{fig2}).

The numbers of the background events from $D^+\to K_L^0\pi^+$ and
$D^+ \to \pi^+\pi^0$, as well as $D^+ \to \tau^+\nu_\tau$, are
estimated by analyzing Monte Carlo samples that are 10 times larger
than the data.  The input branching fractions for $D^+\to K_L^0\pi^+$ and
$D^+ \to \pi^+\pi^0$ are from Ref.~\cite{pdg2010}.
For estimation of the backgrounds from $D^+ \to \tau^+\nu_\tau$ decay, we use
branching fraction $B(D^+ \to \tau^+\nu_\tau)=2.67\times B(D^+ \to \mu^+\nu_\mu)$,
where $B(D^+ \to \mu^+\nu_\mu)$ is quoted from Ref.~\cite{cleo_fd_prd78_p052003_y2008}
and 2.67 is expected by the SM.

The backgrounds from other $D$ decays are corrected considering the
difference in the numbers of events from the data and simulated events
in the range from $0.15$ to $0.60$ GeV$^2$/c$^4$.  Other background
events are from $e^+e^-\to \gamma_{\rm ISR} \psi(3686)$, $e^+e^-\to
\gamma_{\rm ISR} J/\psi$, where $\gamma_{\rm ISR}$ denotes the photon
produced due to initial state radiation, $e^+e^-\to $ q$\bar{\rm q}$
(q = u, d, or s), $e^+e^-\to \tau^+\tau^-$ and $\psi(3770)
\rightarrow$ non-$D\bar D$ decays that satisfy the event-selection
criteria of purely leptonic decays. The numbers of these background
events are estimated by analyzing Monte Carlo samples of each of the
above-listed processes,
which are about 10 times more than the data.
After normalizing these numbers of background events from the Monte Carlo samples to the data,
we expect that there are $42.0 \pm 2.3$ background events,
where the errors reflect the Monte Carlo statistics, uncertainties in the
branching fractions and/or production cross sections for the background channels.

After subtracting the number of background events, $409.0\pm 21.2 \pm
2.3$ signal events ($N^{\rm net}_{\rm sig}$) for $D^+ \rightarrow
\mu^+\nu_{\mu}$ remain, where the first error is statistical and the
second is the systematic associated with the uncertainty of the
background estimate.  The weighted overall efficiency for detecting
${D^+\rightarrow \mu^+\nu_{\mu}}$ decays is determined to be
$\epsilon=0.6403\pm 0.0012$ by analyzing Monte Carlo simulated events
for $D^+\rightarrow \mu^+\nu_{\mu}$ in each tagged $D^-$ mode; here
the error is due to Monte Carlo statistics.  Final state radiation is
included in the Monte Carlo simulation.

Inserting $N_{D^-_{\rm tag}}$, $N^{\rm net}_{\rm sig}$ and $\epsilon$ into
$$B(D^+ \to \mu^+\nu_{\mu})=\frac{N^{\rm net}_{\rm sig}}{N_{D^-_{\rm
tag}}\times \epsilon}$$ and subtracting from the signal a $1.0\%$
contribution coming from $D^+ \to \gamma D^{*+}\to \gamma
\mu^+\nu_{\mu}$~\cite{cleo_fd_prd78_p052003_y2008,
radiative_leptonir_decay_paper}, in which $D^{*+}$ is a virtual vector
or axial-vector meson, yields
$$B(D^+ \to \mu^+\nu_{\mu})=(3.71 \pm 0.19 \pm 0.06)\times 10^{-4}, $$
where the first error is statistical and the second systematic.  This
measured branching fraction is consistent within errors with those
measured at BES-I~\cite{bes1_fD}, BES-II~\cite{bes2_fD}, and
CLEO-c~\cite{cleo_fd_prd78_p052003_y2008}, but with the best
precision.

The systematic uncertainty in the $D^+ \rightarrow \mu^+\nu_{\mu}$
branching fraction determination includes seven contributions: (1) the
uncertainty in the number of $D^-$ tags ($0.5\%$),
which contain the uncertainty in the fit to the $M_{\rm BC}$ distribution ($0.5\%$)
and the difference in the fake $\pi^0$ rates between the data and the Monte
Carlo events ($0.1\%$);  (2) the uncertainty in $\mu$ tracking/identification
($0.1\%/0.8\%$) determined by comparing the $\mu$
tracking/identification efficiencies for data and Monte Carlo events,
where the $\mu^{\pm}$ samples are from the copious $e^+e^- \to
\gamma\mu^+\mu^-$ process; (3) the uncertainty in the $E_{\gamma_{\rm
max}}$ selection requirement ($0.1\%$) determined by comparing doubly
tagged $D\bar D$ hadronic decay events in the data and Monte Carlo; (4)
the uncertainty associated with the choice of the $M^2_{\rm miss}$
signal window ($0.5\%$) determined from changes in the measured
branching fractions using different signal window widths; (5) the
uncertainty in the background estimate ($0.6\%$) due to Monte Carlo
statistics of the simulated backgrounds and uncertainties in the
branching fractions or the production cross sections for the
background channels; (6) the uncertainty in efficiency ($0.2\%$)
arising from the Monte Carlo statistics; (7) the uncertainty in the
radiative correction ($1.0\% $), which we take to be $100\%$ of its
central
value~\cite{cleo_fd_prd78_p052003_y2008,radiative_leptonir_decay_paper}.
The total systematic error determined by adding all the component
errors in quadrature is $1.6\%$.

Inserting the measured branching fraction,
$G_F$,
the mass of the muon,
the mass of the $D^+$ meson
and the lifetime of the $D^+$ meson~\cite{pdg2010}
into Eq.(\ref{eq01}) yields
$$f_{D^+} |V_{\rm cd}| = (45.75\pm1.20\pm0.39)~~\rm MeV,$$
where the first error is statistical and the second systematic arising
mainly from the uncertainties in
the measured branching fraction ($1.6\%$)
and the lifetime of the $D^+$ meson ($0.7\%$)~\cite{pdg2010}.
The total systematic error is $0.9\%$ for $f_{D^+} |V_{\rm cd}|$.

The decay constant $f_{D^+}$ is obtained using as input the CKM matrix
element $|V_{\rm cd}|= 0.22520\pm 0.00065$ from the global fit in the
SM~\cite{pdg2010}. Alternatively, $|V_{\rm cd}|$ is determined using
$f_{D^+}=207\pm 4$ MeV from LQCD~\cite{lqcd_HPQCD_UKQCD} as input.
The results are
$$f_{D^+} = (203.2 \pm 5.3 \pm 1.8)~~~\rm MeV$$
\noindent
and
$$|V_{\rm cd}| = 0.2210 \pm 0.0058 \pm 0.0047,$$
\noindent
where the first errors are statistical and the second systematic arising
mainly from the uncertainties in
the measured branching fraction ($1.6\%$),
the CKM matrix element $|V_{\rm cd}|$ ($0.3 \%$), $f_{D^+}$ ($1.9\%$),
and the lifetime of the $D^+$ meson ($0.7\%$)~\cite{pdg2010}.
The total systematic error is $0.9\%$ for $f_{D^+}$ and $2.1\%$ for $|V_{\rm cd}|$.

Our measured value for $B(D^+ \rightarrow \mu^+\nu_{\mu})$ has the
best precision in the world to date.  The value of $f_{D^+}$ can be
used to validate LQCD calculations of $f_{D^+}$,
thereby producing a
more reliable and precise prediction of $f_{B^+}$.  This $f_{B^+}$
value can in turn be used to improve the precision of the
determination of $|V_{\rm td}|$, and the improved $|V_{\rm cd}|$ and
$|V_{\rm td}|$ can be used for more stringent tests of the SM.

\vspace{0.2cm}

The BESIII Collaboration thanks the staff of BEPCII and the computing center
for their strong support.
This work is supported in part by the Ministry of Science and Technology
of China under Contracts No. 2009CB825204 and 2009CB825200;
National Natural Science Foundation of China (NSFC) under Contracts No. 10625524, 10821063, 10825524, 10835001,
10935007, 11125525, 11235011; Joint Funds of the National Natural Science Foundation of China under
Contracts No. 11079008 and No. 11179007; the Chinese Academy of Sciences (CAS) Large-Scale Scientific Facility Program;
CAS under Contracts No. KJCX2-YW-N29 and No. KJCX2-YW-N45; 100 Talents Program of CAS; German Research Foundation DFG
under Contract No. Collaborative Research Center CRC-1044; Istituto Nazionale di Fisica Nucleare, Italy; Ministry
of Development of Turkey under Contract No. DPT2006K-120470; U. S. Department of Energy under Contracts
No. DE-FG02-04ER41291, No. DE-FG02-05ER41374, No. DE-FG02-94ER40823, and No. DESC0010118; U.S. National Science Foundation;
University of Groningen (RuG) and the Helmholtzzentrum f$\ddot{u}$r Schwerionenforschung GmbH (GSI), Darmstadt;
WCU Program of National Research Foundation of Korea under Contract No. R32-2008-000-10155-0.

\end{document}